# Carrier Density Oscillation in photoexcited Semiconductors

Ebrahim Najafi [1*], Amir Jafari [2], Bolin Liao[3], Ahmed Zewail [1†]

[1] Physical Biology Center for Ultrafast Science and Technology, Arthur Noyes Laboratory of Chemical Physics, California Institute of Technology, Pasadena, CA 91125., USA
Current Address: DuPont Experimental Station, 200 Powder Mill Rd, Wilmington, DE 19808, USA
[*] Corresponding author. E-mail address: enajafi@caltech.com
　　　　　　　　　　　　　　　ebrahim.najafi@chemours.com

[2] Department of Physics and Astronomy, Johns Hopkins University, Baltimore, MD 21218, USA
[3] Department of Mechanical Engineering, University of California, Santa Barbara, Santa Barbara, CA 93106-5070

[†] Deceased

**Abstract**

The perturbation of a semiconductor from the thermodynamic equilibrium often leads to the display of nonlinear dynamics and formation of spatiotemporal patterns due to the spontaneous generation of competing processes. Here, we describe the ultrafast imaging of nonlinear carrier transport in silicon, excited by an intense femtosecond laser pulse. We use scanning ultrafast electron microscopy (SUEM) to show that, at a sufficiently high excitation fluence, the transport of photoexcited carriers slows down by turning into an oscillatory process. We attribute this nonlinear response to the electric field, generated by the spatial separation of these carriers under intrinsic and photo-induced fields; we then provide an advection-diffusion model that mimics the experimental observation. Our finding provides a direct imaging evidence for the electrostatic oscillation of hot carriers in highly excited semiconductors and offers new insights into their spatiotemporal evolution as the equilibrium is recovered.

**Significance Statement**

This work uses time-resolved electron microscopy to study the nonlinear carrier transport on the semiconductor surface at ultrafast timescales. Furthermore, it provides, to our knowledge for the first time, direct "imaging "evidence for the spatiotemporal oscillation of hot carriers, which hinders their long-range transport.  This study elucidates the excitation and relaxation mechanisms in semiconductors at high excitation regimes, an information potentially critical for the design of fast and efficient electronic, thermoelectric and optoelectronic devices.

**Introduction**

The spatiotemporal dynamics of hot carriers play a fundamental role in the performance and efficiency of optoelectronic, thermoelectric and photovoltaic devices (*1, 2*). With the constant supply of energy during the operation of a typical device, carriers are often driven far away from equilibrium where their behavior becomes nonlinear and chaotic; this can lead to current instability and spontaneous pattern formation and ultimately result in information loss or even catastrophic failure (*3*). To prevent such an adverse response, it is essential to examine hot carriers in space and time to understand their relaxation pathways as well as their interactions with geometrical complexities such as surfaces, interfaces and defects.

Carrier dynamics are generally characterized by optical pump-probe spectroscopy, which conventionally measures the transient reflectivity or transmission of samples following the excitation by an optical pulse. However, the spatial resolution of this technique is inherently limited by diffraction (*4*); in addition, the large penetration depth of the optical pulse generates a strong contribution from the bulk or the substrate that dominates the overall signal. This potentially leads to misleading conclusions when analyzing samples with reduced dimensions such as thin films and nanomaterials. With the recent development of scanning ultrafast electron microscopy (SUEM), which combines the temporal resolution of the femtosecond laser with the spatial resolution and the surface sensitivity of the electron probe (*5-9*), we are now able to overcome these limitations by "directly" imaging carriers and mapping their evolution in space and time.

The spatiotemporal examination of carriers is essential to reveal their transport and relaxation channels in semiconductors. For instance, Kumar et al. have recently reported a super-ballistic transport of electrons through a geometrical constrain in a graphene device (*10*), which was previously predicted as a means to promote low-loss currents even at elevated temperatures (*11*). Similarly, we recently provided experimental evidence for anisotropic carrier diffusion in black phosphorus, which showed significantly faster in-plane diffusion of holes along the armchair direction than the zigzag direction (*12*).

Here, we study the transport of hot carriers in silicon, excited by an intense ultrashort laser pulse. In this regime, we report a nonlinear oscillatory response as revealed by the close examination of SUEM images. To our knowledge, this is the first "microscopy" evidence of hot carrier oscillation in semiconductors imaged at ultrafast timescales. Furthermore, we report a dramatic reduction in the transport rate relative to lower fluences. We attribute this behavior to a transient electric field generated by the spatial separation of the photoexcited electrons and holes, which strongly opposes their diffusion. The insight provided here enriches our current understanding of nonlinear processes in highly perturbed semiconductors; this knowledge is essential for the development of the next generation devices that routinely operate under extreme excitations.

In this work, we used a p-type silicon wafer with the doping concentrations of $2.1 \times 10^{19}$ B cm$^{-3}$, which was purchased from MTI Corp. and used without any modification. The doping type and concentration were further confirmed by energy dispersive X-ray analysis. The sample was cleaved and immediately transferred into the SUEM chamber where the vacuum was maintained at $1.2 \times 10^{-6}$ torr. To record the SUEM images, the sample was excited by a Gaussian laser pulse with the wavelength of 515 nm and the fluence of 3.0 mJ cm$^{-2}$ at 2 MHz repetition rate. This fluence was well below the damage threshold of silicon (*13*); the examination of the surface after the measurement showed no significant damage was produced on the surface. A brief description of SUEM is summarized here, and discussed extensively elsewhere (*7, 14-15*). In SUEM, femtosecond infrared pulses (1,030 nm wavelength, 300 fs duration) were generated by a Clark-MXR fiber laser system and used to produce green (515 nm) and UV (257 nm) pulses. The spatial intensity profile of the laser was measured at a point perpendicular to the direction of propagation to ensure a gaussian laser beam. The green pulse excites the specimen while the UV pulse is directed toward the photocathode where it emits a short electron pulse from the field emission tip (Figure 1). The time delay between the pump and the probe pulses is controlled by a mechanical delay stage, which covers a range between −760 ps to 3.4 ns and provides a snapshot resolution of 1 ps. The temporal resolution of the electron pulse was measured from the optical excitation of intrinsic silicon and determined to be better than 1 ps. The electron pulse, accelerated to 30.0 kV, raster-scans the surface, generating secondary electrons (SEs) which are then collected by an Everhart–Thornley detector and translated into pixel intensities to form images. To obtain dynamic images, a series of SUEM images were captured after optical excitation. In order to avoid systematic errors in the measurement, these images were acquired randomly at predetermined times. A reference image, an average of several images recorded prior to optical excitation, was then subtracted from these images to remove the background as well as to enhance the dynamic signal. Since the background in raw images contains elemental and topographic information of the surface, its removal ensures that the remaining signal is merely due to the laser-induced changes in carrier density. Consequently, in the resulting "contrast images", the bright and dark signals are interpreted as increases in the local free electron and hole densities, respectively. Further explanation of the contrast mechanism is presented elsewhere *(16)*. In short, the emergence of a contrast in doped silicon is due to the changes in the surface potential, driven by the transports of the minority carriers toward the surface after optical excitation. This impacts the SE emission intensity by altering the effective electron affinity at the surface, and results in a bright contrast in p-type and a dark contrast in n-type silicon. Typically, SUEM provides better than 10 nm spatial resolution, ideal to study nanomaterials; however, for the current measurement, 200 nm resolution was sufficient to resolve the spatial dynamics at the surface. While the volume excited by the electron pulse can extend deep into the sample, the SEs are only emitted from the top 5−10 nm below the surface, thus providing surface sensitive information.

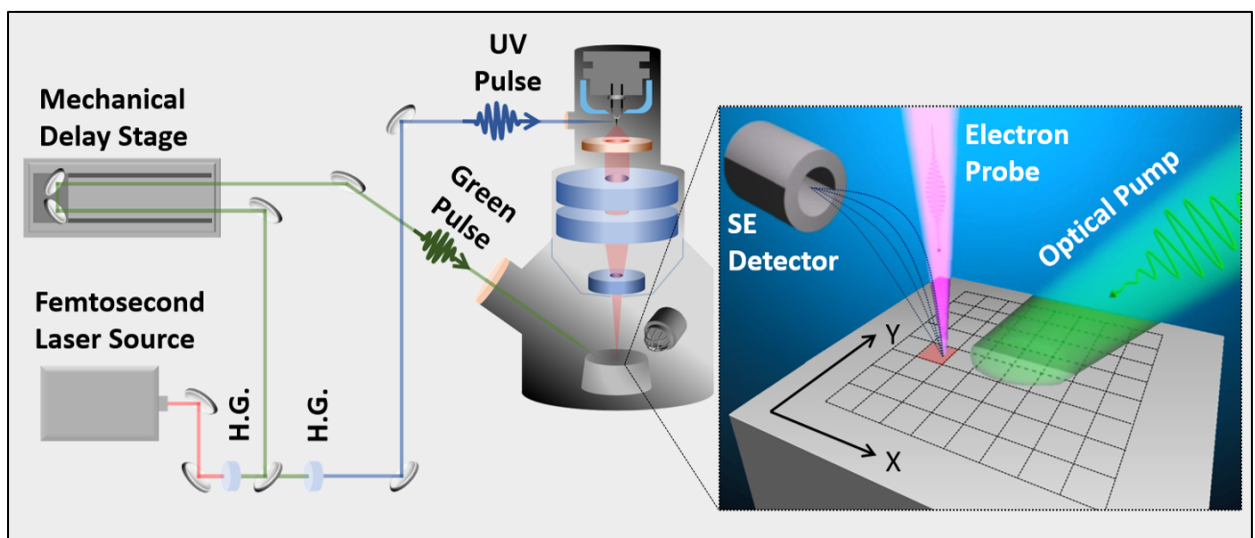

**Figure 1-** Schematic representation of the SUEM setup. In SUEM, a femtosecond fiber laser produces IR pulses (1,030 nm wavelength, 300 fs duration, 2 MHz repetition rate), which are split and converted into green and UV pulses by harmonic generation. The green excites the sample and the UV generates a short electron pulse at the tip, which raster scans the surface and measure its spatiotemporal response to the optical excitation. The time delay between the green pump and the electron probe is controlled by a mechanical delay stage, which provides 1 ps resolution between -760 ps and 3.4 ns.

**Manuscript Text**

The SUEM images, presented in Figure 2A, demonstrate the spatiotemporal evolution of the excess carriers as the photoexcited silicon evolves toward the ground state. The sample shows no observable contrast in the negative time, indicative of its return to the ground state between two consecutive pulses separated by 500 ns. The optical excitation of the sample results in a large density of carriers, as is evident by the emergence of a strong bright contrast at the crossover whose intensity systematically increases over time and maximizes by 50 ps. The carrier population then expands laterally, giving rise to broader spatial profiles. The expansion of the population distribution appears much faster below 500 ps, but at extended times it transitions into a much slower process. Furthermore, a closer look at the images shows an alternating pattern emerging away from the center of the excitation by 150 ps, which appears to gradually decay at extended times. We ensured that this pattern is real and not artifacts formed by a misshaped laser beam or the interference between the beam and misaligned optical components in the beamline.

To obtain in-depth and quantitative information from these contrast images, we extracted the spatiotemporal profiles along the dotted line in Figure 2A and plotted in Figure 2B. In the negative time, the profile is effectively flat due to the absence of photoexcited carriers. By 50 ps after optical excitation, a uniform Gaussian peak develops which resembles the spatial geometry of the laser at the crossover and reflects the distribution of the photoexcited carriers on the silicon surface. The diffusion of carriers broadens the peak by 77 ps, reducing the carrier density at the center of the profile while increasing near the rims as evident by the formation of two side-peaks. While there is a slight reduction in the width of the distribution at 90 ps, the overall trend shows further broadening at extended times; however, at approximately 350 ps, carrier transport direction appears to temporarily reverse in order to partially restore the initial spatial distribution. At extended times, the distribution gradually broadens, which is consistent with the normal diffusion.

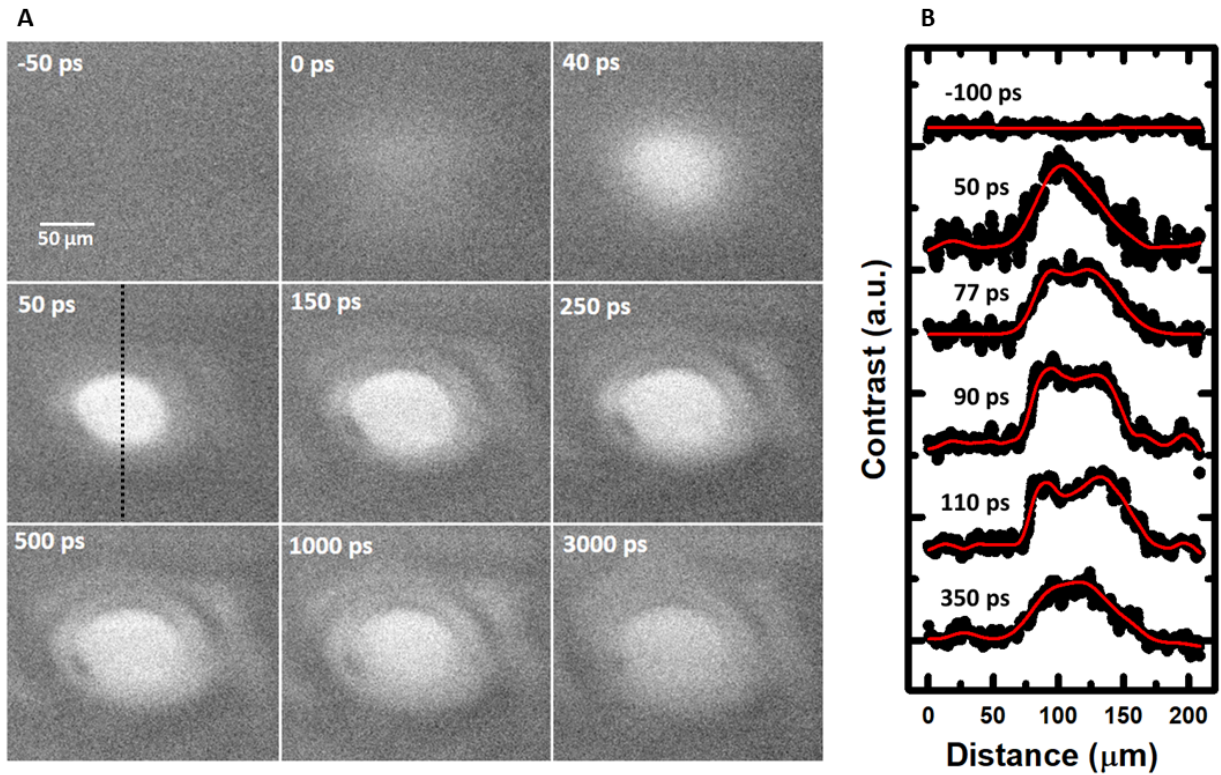

**Figure 2-** **(A)** Selected SUEM images showing the dynamics of photoexcited carriers in the silicon sample. The images show the unperturbed surface ($t \ll 0\,ps$) as well as the excitation ($t \cong 0\,ps$), ultrafast expansion (50−500 ps), and steady-state diffusion ($t > 500\,ps$) of photoexcited carriers. Furthermore, by 150 ps, an alternating pattern emerges away from the crossover which decays at later times. **(B)** The spatial profile extracted along the dotted line and plotted for several time delays; the red lines are to guide the eye.

In order to gain further insights into the spatial dynamics of photoexcited carriers, we determine their transient second moment, $\sigma^2$, by analyzing the contrast images:

$$\sigma^2(t) = \frac{\iint (x^2 + y^2) I(x,y) dxdy}{\iint I(x,y) dxdy} \quad (1)$$

where $I(x,y)$ is the contrast intensity of the pixel located at the coordinates $(x,y)$ for the snapshot recorded at time $t$. The center of the distribution is determined from the data and is taken to be the origin of the $x$ and $y$ axes in equation (1).

We first begin by analyzing $\sigma^2(t)$ of the sample excited at a much lower fluence, 0.16 mJ cm$^{-2}$, as shown in Figure 3A. The graph clearly shows two distinct transport regimes, an initial ultrafast expansion observed at early times, which causes the distribution to widen by ~25% and a normal diffusion beyond ~500 ps. We previously attributed this behavior to the initial high temperature in the excited volume which exponentially reduces as carriers diffuse *(17)*. While we observe a similar fast expansion and a subsequent normal diffusion for the sample excited at 3.0 mJ cm$^{-2}$ (Figure 3B), there are dramatic differences between the graphs that deserve further exploration. For instance, the extent of the spatial expansion is approximately ~15%, which is smaller than the former despite the significantly higher laser fluence. Furthermore, prior to transitioning into a normal diffusion, $\sigma^2(t)$ presents a few oscillations, which become stronger as carriers transport further away.

At 0.16 mJ cm$^{-2}$, where the transport is approximated merely by diffusion, $\sigma^2(t)$ determines the initial diffusion coefficient to be $D\sim10^4$ cm$^2$ s$^{-1}$ and this diffusivity increases monotonically with the laser fluence *(17)*. Generally, in a normal diffusion process, the second moment and diffusivity are related by $\sigma^2(t) = 4Dt$. Here, $\sigma^2(t)$ is convoluted by a complex combination of transport processes, which prevents the normal diffusion approximation. Thus, we define the transport rate $v = d\sigma^2/dt$ for these two excitation regimes and as shown in Figure 4 to evaluate how photoexcited carriers transport in space and time. At 0.16 mJ cm$^{-2}$, transport rate is $v\sim10^4$ cm$^2$ s$^{-1}$, which exponentially decreases to $v\sim20$ cm$^2$ s$^{-1}$ as carriers lose their excess energy to the lattice (blue circles). This is consistent with the diffusive transport approximation at a low laser fluence *(17)*.

At 3.0 mJ cm$^{-2}$, the initial transport rate is ~10 times slower than the former while still decaying exponentially at extended times (red circles). This transport rate periodically oscillates, indicating that there is an underlying mechanism that impacts free carrier diffusion. Ultimately at extended times, this periodic oscillation disappears as the normal diffusion takes over.

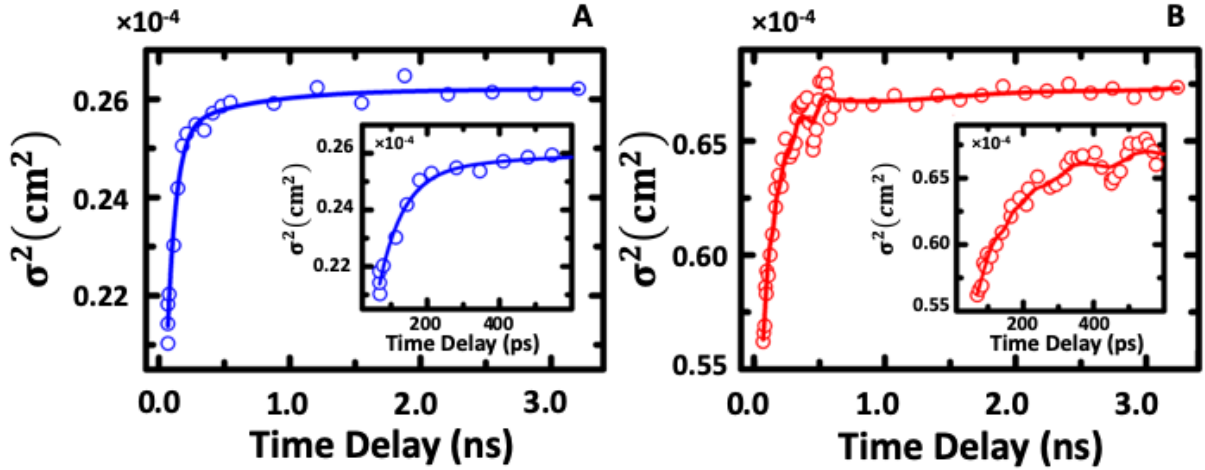

**Figure 3-** Time evolution of the second moment $\sigma^2$ of the excited carrier distribution in p-type silicon excited at 0.16 mJ cm$^{-2}$ (A) and 3.0 mJ cm$^{-2}$ (B). The circles are experimental data, and the line are to guide the eye. Figure 3B clearly shows subtle oscillations below 600 ps (inset), which are absent in Figure 3A (inset).

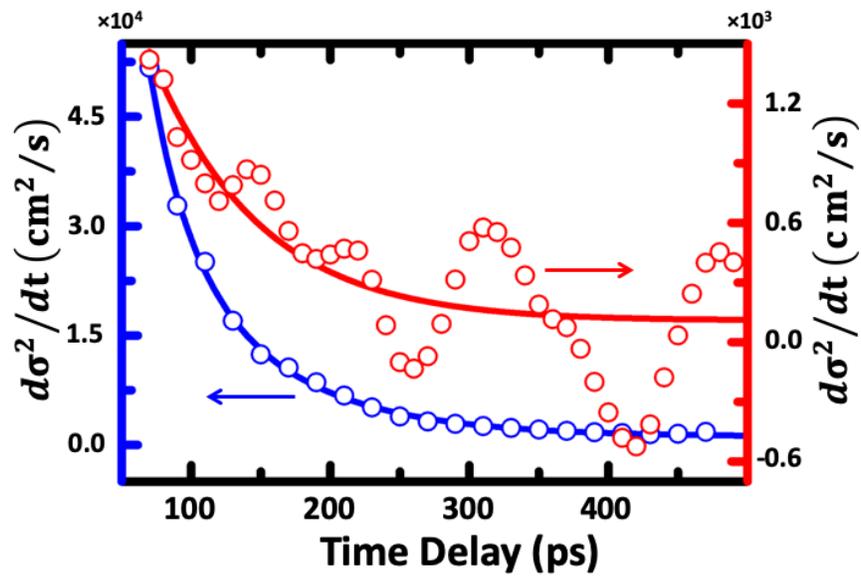

**Figure 4-** Transport rate ($v = d\sigma^2/dt$) of the excited carrier distribution in the silicon sample excited at $0.16$ mJ cm$^{-2}$ (blue circles) and $3.0$ mJ cm$^{-2}$ (red circles). The solid lines represent the exponential nature of the decay over time.

We have developed a theoretical model that accounts for this unusual observation by employing the diffusion of photoexcited carriers as well as the electric field (**E**) generated by their separation in space and time. We expect the optical excitation of silicon with the laser pulse to generate large densities of free electrons and holes across the bandgap with more than 1 eV excess energy. The band bending in the space-charge region (SCR) in the silicon subsurface spatially separates these carriers by transporting electrons toward the surface and holes into the bulk; in a highly doped p-type silicon, the SCR extends only a few nanometers into the bulk *(18-20)*. Once separated by the SCR, the large pressure build-up within the excited region results in their fast diffusion (*18*); since holes travel at slower velocities than electrons due to their higher effective mass, a net positive is generated at the center of the distribution. This induces a dynamic electric field whose strength depends on the extent of the electron-hole separation (Figure 5). This field the diffusive transport as soon as carriers are separated, but its restoring effect becomes visible only when it is large enough to slow down electrons. In practice, however, the initial fast diffusion weakens quickly as carriers lose their excess energy to the lattice through carrier-phonon coupling; similarly, the electrostatic interaction between the electron and hole densities reduces when carrier recombination dominates and decreases the number of free carriers.

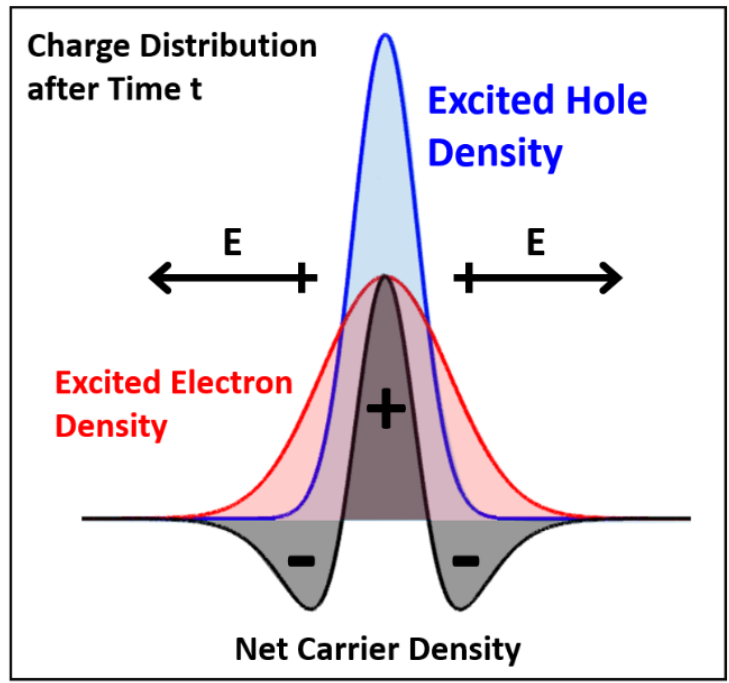

**Figure 5-** Schematics of photoexcited carrier transport in silicon. Starting with Gaussian distributions, electrons travel faster than holes due to their lower effective mass. This results in a net positive distribution at the center and a net negative distribution around the perimeter, effectively creating a dynamic dipole that hinders diffusion.

To formulate this model in the cylindrical coordinates, we begin with the excitation of the sample with a laser pulse that generates the initial carrier distribution $(n_{e,h})$:

$$n_{e,h}(r,z) = \left(\frac{1}{2\pi\sigma_{e,h}^2} e^{-r^2/2\sigma_{e,h}^2}\right) \frac{\alpha(1-R)}{E} P_0 e^{-z\alpha}, \qquad (2)$$

where $\sigma_{e,h}^2$ is the second moment of the electron and hole distributions, $\alpha$ is the absorption coefficient, $R$ is the reflectivity, $P_0$ is the laser intensity at the surface and $E$ is the photon energy. Using the experimental absorption of intrinsic silicon at 2.4 eV incident photon energy, we estimate the excited carrier concentrations to be $\sim 2.0\times 10^{20}$ cm$^{-3}$. In practice, we expect stronger excitation due to bandgap narrowing in our sample by tens of meV (21). For the laser power density used here, two-photon absorption processes in silicon are negligible as they make up only a small fraction of the optical transitions (22). The average initial excited carrier temperature induced by the photon energy E=2.4 eV is estimated to be $T_0^*(0) = E/k_B \sim 28{,}000$ K in our conditions; therefore, the initial carrier average temperature far exceeds the Fermi temperature and leads to the formation of a non-degenerate carrier gas. Gitomer et al. have previously measured the temperature profile of hot carriers under intense laser illumination from their bremsstrahlung X-ray spectrum (23). For the power density employed in this study, they found the highest carrier energy to be 30 eV, corresponding to $T_0^*(0) = 3.5\times 10^5$ K. Because SUEM measures the SEs emitting from the sample, it is highly sensitive to the carrier energy; thus, more energetic carriers contribute more to the SE emission. Therefore, we preferentially observe the diffusion of carriers in the high-energy tail of the excited carrier distribution. Accordingly, we expect to observe effective carrier temperatures in the $\sim 1$-$5\times 10^5$ K range.

In the continuity equation, $\partial_t n_{e,h}(\mathbf{x},t) = -\nabla \cdot \mathbf{J}_{e,h}(\mathbf{x},t)$, flux is related to carrier diffusion and advection as $\mathbf{J}_{e,h} = D_{e,h}\nabla n_{e,h} + n_{e,h} v_{e,h}$, where $D_{e,h}$ is the diffusivity and $v_{e,h}$ is the advection velocity. Since the total electric field is given by $\mathbf{E}_{Total} = \mathbf{E}_{SCR} - \nabla\phi$, where $\mathbf{E}_{SCR}$ is the electric field in the space-charge region in the subsurface, the advection-diffusion equation becomes:

$$\frac{\partial n_{e,h}(\mathbf{x},t)}{\partial t} = \nabla \cdot \left(D_{e,h}\nabla n_{e,h}(\mathbf{x},t) + \mu_{e,h}\nabla n_{e,h}(\mathbf{x},t)[\mathbf{E}_{SCR} - \nabla\phi(\mathbf{x},t)]\right) \qquad (3)$$

Note that $\phi$ is obtained from the Poisson's equation, which relates $\phi(\mathbf{x},t)$ to the sum of the electric scalar potentials of electrons $(\phi_e)$ and holes $(\phi_h)$:

$$\nabla^2 \phi(\mathbf{x},t) = +e[n_h(\mathbf{x},t) - n_e(\mathbf{x},t)] \qquad (4)$$

The boundary conditions for the Poisson's equation are given by:

$$\begin{cases} \phi(x = \pm a, y, t) = 0 \\ \phi(x, y = \pm a, t) = 0 \end{cases} \qquad (5)$$

where $a$ is the "box size" taken large enough to approximate the boundaries at infinity. The effective temperature $(T^*)$ of the excited carriers decays exponentially with a characteristic relaxation time $\tau$ *(17)*:

$$T^*(t) = T^*(0)e^{-t/\tau} + T_0 \qquad (6)$$

where $T_0$ is the room temperature. From the Einstein relations for transport equation, we write $D_{e,h}(t) = \mu_{e,h} k_B T(t)/e$ where $k_B$ is Boltzmann's constant and $\mu_{e,h}$ is the mobility of carriers, which strongly depends on the electric field *(24)*:

$$\mu_i(E) = \begin{cases} \mu_i(0), & E \sim 0 \\ \mu_i(0)\left(1 + \dfrac{\sqrt{|-\nabla\phi|^2 + E_D^2}}{E_C}\right)^{-1/2}, & E \gg E_C \end{cases} \qquad (7)$$

Here, $E_C$ is the critical electric field, which is estimated to be $7.5 \times 10^5$ $(V/m)$ for p-type silicon. Thus, equation 4 becomes:

$$\frac{\partial n_{e,h}(\mathbf{x}, t)}{\partial t} = \left(\frac{\mu_{e,h} k_B}{e} T^*(0)e^{-t/\tau} + \frac{\mu_{e,h} k_B}{e} T_0\right) \Delta n(X, t) + \nabla \cdot \left(\mu_{e,h} n(\mathbf{x}, t) \nabla \phi(\mathbf{x}, t)\right) \qquad (8)$$

We then numerically solved these coupled equations using finite element method for a system similar to that used in the experiment using zero flux boundary condition. Specifically, we employed a three-dimensional geometry with 0.5 V potential across the SCR in the subsurface and a Gaussian laser pulse (25×25μm²) that exponentially decayed into the bulk.

Figure 6A plots the spatiotemporal evolution of the net carriers $(n_e - n_h)$ on the surface. As previously discussed *(16)*, $n_e - n_h$ is an approximation of SUEM contrast since it reflects the transient changes in the surface potential in doped silicon. Initially, the net carrier intensity is flat since the starting electron and hole densities are equal. Photoexcited electrons then partly transport toward the surface while holes drift into the bulk due to the built-in electric field in the SCR. Simultaneously, these free carriers undergo fast diffusion due to their high temperature; since electrons diffuse faster than holes, due to their lighter effective mass, they concentrate at the edges of the distribution, leaving behind a positive center at extended times. At 300 ps, however, the intensity at the center appears to regrow due to the partial repopulation by electrons. This is in a relatively good agreement with the experimental observation in Figure 2 and corresponds to negative $v$ as seen in Figure 4. This model also accounts for the formation of the ring-like structure as evident by the development of subtle peaks at the shoulders, which emerge from the superposition of the separated electrons and holes across the sample thickness and normal to the surface. In fact, when the simulation is repeated for the a two-dimensional geometry, this structure completely disappears.

Figure 6B plots the simulated second moment of the carrier distribution versus time. The graph clearly shows initial fast diffusion that decays exponentially into a room temperature diffusion after a few hundreds of picoseconds. The overall slow diffusivity in the simulation is consistent with the experimental observation at such a high excitation density, in agreement with the resisting effect of the electric field on carrier diffusion as electron and hole distributions separate in the space and time. Furthermore, the simulation repeated at different carrier densities suggests that this electric field, though present at all excitation densities, is strong enough to interfere with the diffusion only at high excitation densities. The subtle oscillations in the initial transport also reflects the electrostatic interactions between the electron and hole densities as they separate in space and time (inset, Figure 6A ). However, the magnitude of the oscillations is smaller than we have observed experimentally in Figure 3B, which is probably due to the exclusion of a few fundamental factors such as carrier recombination for the sake of simplicity.

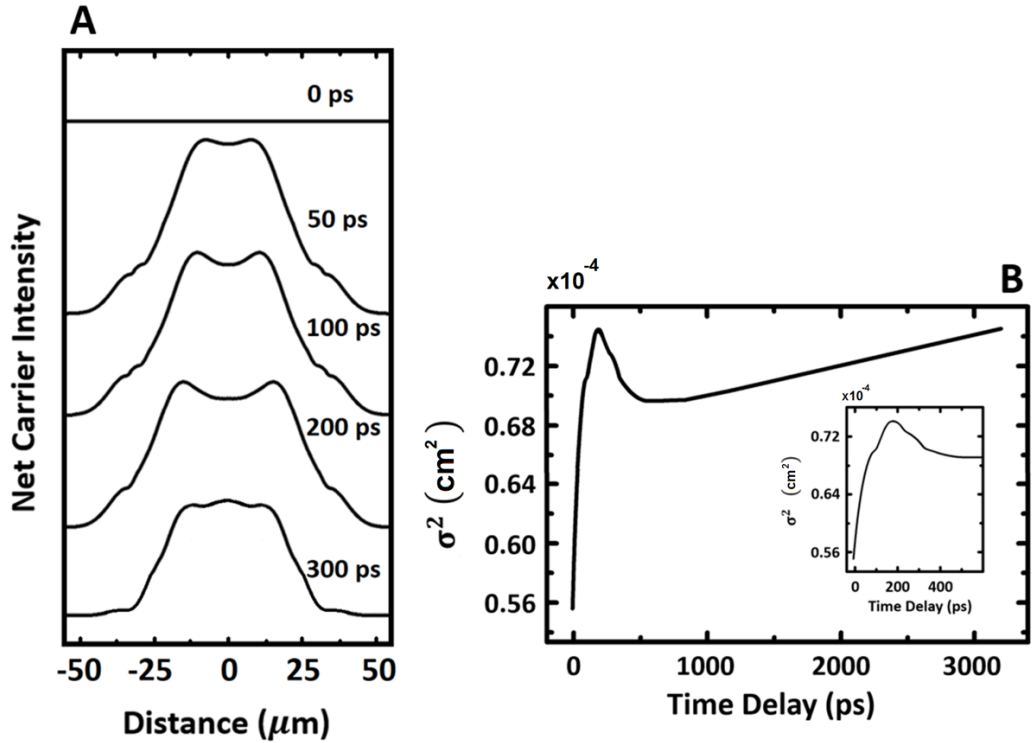

**Figure 6**- Numerical simulation of hot carrier transport on the silicon surface. (A) is the spatiotemporal evolution of carriers across the photoexcited region. (B) The second moment of the carrier distribution plotted versus time, which shows their fast and normal diffusion; the oscillatory nature of the transport is manifested as momentary changes in the second moment at early times as shown in the inset.

In conclusion, we used four-dimensional electron microscopy to study the spatial dynamics of carriers on the silicon surface at a high excitation fluence. We showed that the transport of carriers in this regime was strongly influenced by the electrostatic interactions induced by the spatial separation of the photoexcited electron and hole densities. This study reveals the ultrafast behavior of hot carriers in highly perturbed systems, which can potentially lead to the design of more efficient devices with minimal loss of energy and information.

**Acknowledgements**

This work was supported by NSF grant DMR-0964886 and Air Force Office of Scientific Research Grant FA9550-11-1-0055 in the Physical Biology Center for Ultrafast Science and Technology at California Institute of Technology, which is supported by the Gordon and Betty Moore Foundation.